**Memristor Crossbars with 4.5 Terabits-per-Inch-Square Density and 2 nm Dimension**

Shuang Pi[1], Can Li[1], Hao Jiang[1], Weiwei Xia[2], Huolin Xin[2], J. Joshua Yang[1], Qiangfei Xia[1*]

[1] Department of Electrical and Computer Engineering, University of Massachusetts, Amherst, Massachusetts 01003, USA

[2] Center for Functional Nanomaterials, Brookhaven National Laboratory, Upton, New York 11973, USA

* Email: qxia@umass.edu

**Memristor is a promising building block for the next generation nonvolatile random access memory and bio-inspired computing systems. Organizing memristors into high density crossbar arrays, although challenging, is critical to meet the ever-growing high capacity and low energy demands of these applications especially in the big data era. Here, we construct memristor crossbars with a single-layer density up to 4.5 terabits per inch square, an order of magnitude denser than the state-of-the-art 64-layer triple level cell NAND flash technology. The memristors in the crossbars are $2 \times 2$ nm$^2$ in size, capable of switching with tens of nano ampere electric current. The densely packed memristor crossbars of extremely small working devices provides a power-efficient solution for high density information storage and processing.**

Memristors uses tunable non-volatile resistance to represent digital or analogue information in both memory and computing applications [1-10]. Owing to its simplicity in structure, scalability is a key advantage of memristor over other emerging electronic devices. Discrete devices at sub-10 nm size or random accessible working arrays with a packing density up to 64.5 Gbit/in$^2$ have been demonstrated in previous studies [11-15]. While even smaller devices and higher packing density are required in order to meet the ever-increasing speed-energy efficiency demand in applications [4,9], both are difficult to achieve because of challenges in making low-resistance nanoelectrodes and packing them into high-density random accessible crossbars [12,14]. Although conductive nanomaterials such as carbon nanotube, graphene nanoribbons and "interface-free" dopant wires have been developed [16,17], it is challenging to efficiently pack these nanomaterials into dense and ordered electrode arrays [18-21]. While alternative techniques such as super-lattice nanowire pattern transfer (SNAP) have been proven to be an effective approach for generating functional electrodes on an adhesive thin film with 33 nm pitch [22,23], it remains to be demonstrated whether they are suitable for making fully random accessible arrays. In addition, all the aforementioned methods have encountered dramatically increased resistance of the electrodes at nanoscale, imposing significant barrier for their adoption in operational circuits [17,24-26]. Furthermore, the write/read current of an individual memristor in a densely packed array could lead to crosstalks that limit parallel operation of the array [27,28].

Herein we explore the extreme memristor scaling and demonstrate functional crossbar arrays of $2 \times 2$ nm$^2$ memristor with 4.5 Tbit/inch$^2$ packing density for the first time. We introduce metal nanostructures of ultrahigh height/width aspect ratio (up to 1500), which we refer to as nanofins, as the electrodes. The nanofins are fabricated by sidewall deposition of metal (e.g., platinum) and dielectric (e.g., alumina) thin films, followed by chemical mechanical polishing (CMP) to expose the metal arrays in a planarized plane. Single metallic structures with lower aspect ratios were made through sidewall deposition earlier but they were never organized into ultra-dense random-accessible arrays [13,29,30]. Through proper materials engineering, we achieved continuous ultrathin platinum nanofins at 2 nm thickness with a resistance well below 100 $\Omega$/µm, 10 times less resistive than a 10 nm wide Cu wire made with the most advanced complementary metal oxide semiconductor (CMOS) technology [26,31]. Crossbars of memristors are made by stacking two nanofin arrays with a thin layer of hafnium oxide/titanium oxide sandwiched in between. The $2 \times 2$ nm$^2$ memristors in the entire array can be independently switched with low current (< 50 nA), suggesting a great potential of these devices for power-efficient data storage and unconventional computing applications.

Figure 1a schematically illustrates the nanofin structure. Compared with a conventional nanowire, the aspect ratio is significantly higher. Consequently, the resistance of the nanofin along the length direction is much smaller. For example, a 2 nm wide Pt nanofin with an aspect ratio of 1500 was measured to be 65 $\Omega$/µm. This is over 1000 times more conductive than a Cu nanowire with projected dimension of 2 nm width and 2 nm height, and four orders of magnitude less resistive than a carbon nanotube of similar diameter (Fig. 1b) [26,32]. Figure 1c shows the key steps of nanofin fabrication and memristor array integration. First, the 2 nm Pt thin film with a 2 nm Ge wetting layer is deposited onto the side wall of a $SiO_2$ trench, followed by the patterning and deposition of two Cu contact pads and an alumina isolation layer. This procedure is repeated a number of times to generate a multi-fin array. The layer-by-layer fabrication fashion allows for the construction of contact pads to address each electrode and hence makes the final array randomly accessible. The chip is then passivated with chemical vapor deposited (CVD) $SiO_2$ and planarized by chemical-mechanical polishing (CMP), resulting in a smooth surface with the top edges of nanofins exposed. Finally direct wafer bonding is applied to bring two polished chips face to face, with a thin layer of hafnium oxide/titanium oxide sandwiched in between as the switching medium. A more detailed description of the fabrication procedure is illustrated in *Supplementary Figs. 1-8*.

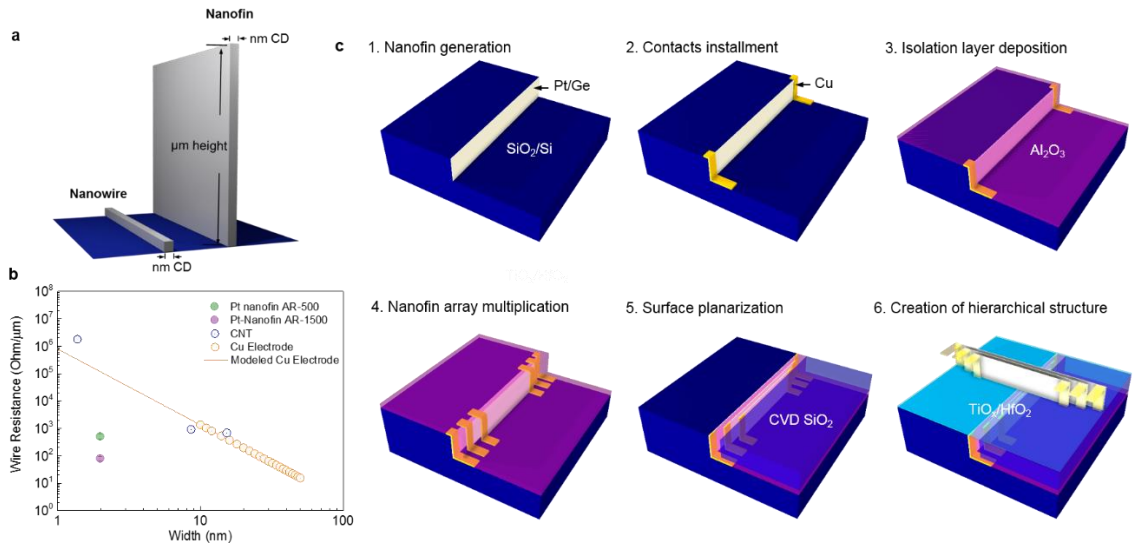

**Figure 1 | Nanofin enables low resistance electrodes for memristor crossbars. a**. Schematic of a nanofin structure, drawn side by side with a conventional nanowire to highlight the high aspect ratio. **b**. Comparison of the resistance of nanofins with other nanostructures. With an aspect ratio of 1500 to 1, the 2 nm Pt nanofin is 10 times more conductive than a Cu wire of 10 nm, and over four orders of magnitude more conductive than a CNT of 1.38 nm diameter. The Cu electrode and CNT data are from refs. 26,32. **c**. Process flow of constructing memristor crossbars with nanofin electrode. It involves the repeated deposition of metal thin film, contact and isolation layer on the sidewall of a silicon oxide trench (1-4), planarization and polishing to expose the metal edges (5) and bonding of two nanofin chips with switching layer sandwiched in between (6). A step-by-step description is presented in the supplementary information.

One of the biggest challenges in the array fabrication is to achieve high quality nanofins, because the high surface energy of noble metals (e.g., Cu, Pt) favors the formation of isolated islands or percolated networks rather than continuous ultrathin metal films on an oxide surface [33]. We solved this problem by inserting a 2 nm Ge thin film between Pt and $SiO_2$, which significantly improved the wettability of Pt and hence the thin film quality. The root mean square (RMS) surface roughness of 2 nm Pt deposited directly on a $SiO_2$ surface was measured to be 0.503 nm, while that with a 2 nm Ge wetting layer was 0.123 nm, close to that of a high quality thermal $SiO_2$ surface (Figs. 2a-c). The RMS roughness of Pt/Ge thin films remained at ~0.1 nm regardless of the Pt thickness till it reaches sub-1 nm domain (*Supplementary Figs. 9 –11*). This finding suggests that the scaling limit of Pt nanofins, with the help of the Ge wetting layer, is around 1 nm, below which only discontinuous membrane could be formed. Figures. 2d and 2e show typical cross sectional transmission electron microscopic (TEM) images of nanofin arrays with 2 and 1.5 nm linewidth, respectively. These nanofins are up to 680 µm long, 1 - 3 µm high, and are all continuous with 100% fabrication yield.

To electrically isolate each nanofin in the array, we deposited 8 nm $Al_2O_3$ by atomic layer deposition (ALD) in between the metal fins. We chose alumina because of its high dielectric strength (7 MV/cm) that suppresses leakage current and avoids shortage between neighboring electrodes when biased at different voltages. The fabricated nanofins (2 nm width, 1 µm height) were electrically characterized through the Cu contact pads (Fig. 3a). We confirmed that each metal nanofin is electrically isolated from each other as the leakage current is lower than 1 pA (Fig. 3b). Furthermore, the resistances of 200, 440 and 680 µm long nanofins in the same array were measured to be 338, 175, 108 × $10^3$ Ω, respectively, which are equivalent to an average wire resistance of 470 Ω/µm (Fig. 1b). The well isolated and highly conductive nanofins are suitable for building high density memristor crossbars with negligible crosstalk.

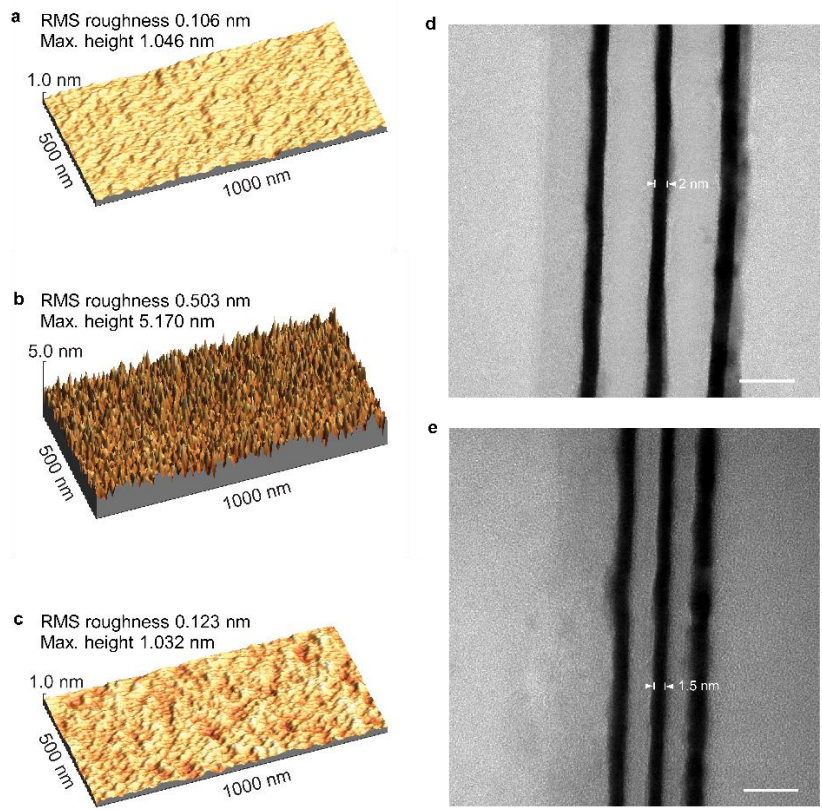

**Figure 2 | Continuous Pt nanofin arrays fabricated on SiO$_2$ with Ge wetting layers. a**. Atomic force microscopic (AFM) image of a thermal SiO$_2$ on Si substrate with RMS roughness of 0.106 nm. **b**. AFM image of 2 nm Pt thin film prepared by directly evaporation of Pt on the oxide substrate. The RMS roughness is 0.503 nm. AFM image of 2 nm Pt thin film evaporated on the oxide substrate with a 2 nm Ge wetting layer. **c**. The RMS roughness is 0.123 nm, indicating the formation of continuous Pt thin film. **d** and **e** are cross sectional TEM image of nanofin arrays with 2 nm width/12 nm pitch and 1.5 nm width/7 nm pitch, respectively. Scale bars: 10 nm.

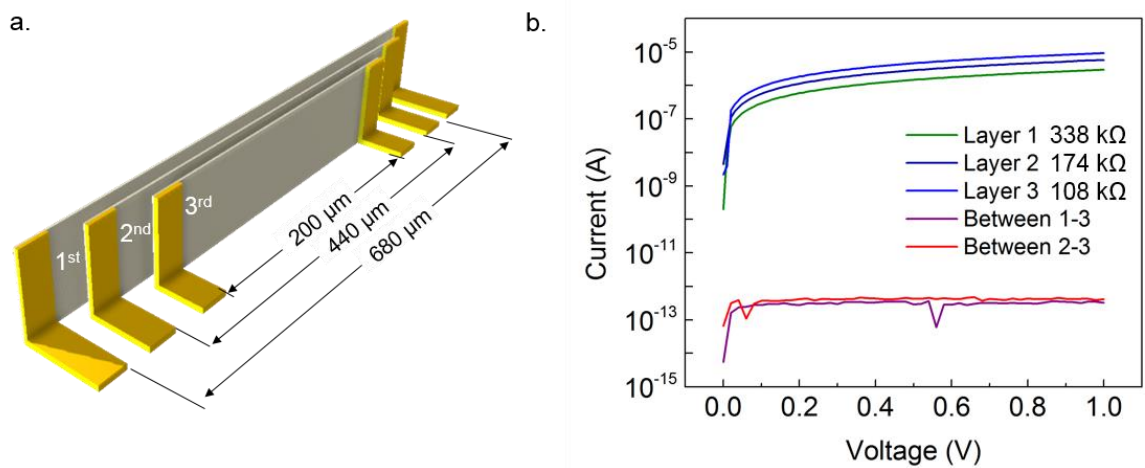

**Figure 3 | Highly conductive and well isolated 2 nm nanofin electrode array. a**. Schematic illustration of three nanofins in an array, each connected to two copper contact pads through which random access is made possible. **b**. I-V characterization of the electrode array, which shows that the nanofins are electrically isolated from each other, while each individual one is very conductive. The conduction current of every nanofin electrode is plotted in green, royal blue and blue. The leakage currents between neighboring nanofins are plotted in purple and magenta.

To make memristor crossbar, we deposited a 7 nm switching layer (3 nm $TiO_x$ plus 4 nm $HfO_2$) on one polished nanofin chip, and bonded it to another chip with the two nanofin arrays orthogonal to each other. Figure 4a shows a representative TEM image of a 3 × 3 array (see *Supplementary Information* for the TEM sample preparation). The junction area of each cell is 2 × 2 $nm^2$ and the pitch is 12 nm, corresponding to a 4.5 Tbit/$inch^2$ packing density (see *Supplementary Fig. 14* for the precise measurement). This is, to the best of our knowledge, the first electronic circuits with individual components scaled down to 2 nm dimension. Every cross point is composed of a pair of perpendicular Pt nanofins and the 7 nm thick switching medium sandwiched in between. Further elementary analyses on the array using energy-dispersive X-ray spectroscopy (EDX) mapping confirmed the constitutions of nanofin electrodes and the crossbar networks. Highly ordered electrode structures with well patterned isolation layers are evidently displayed (Fig. 4b and *Supplementary Fig. 15*).

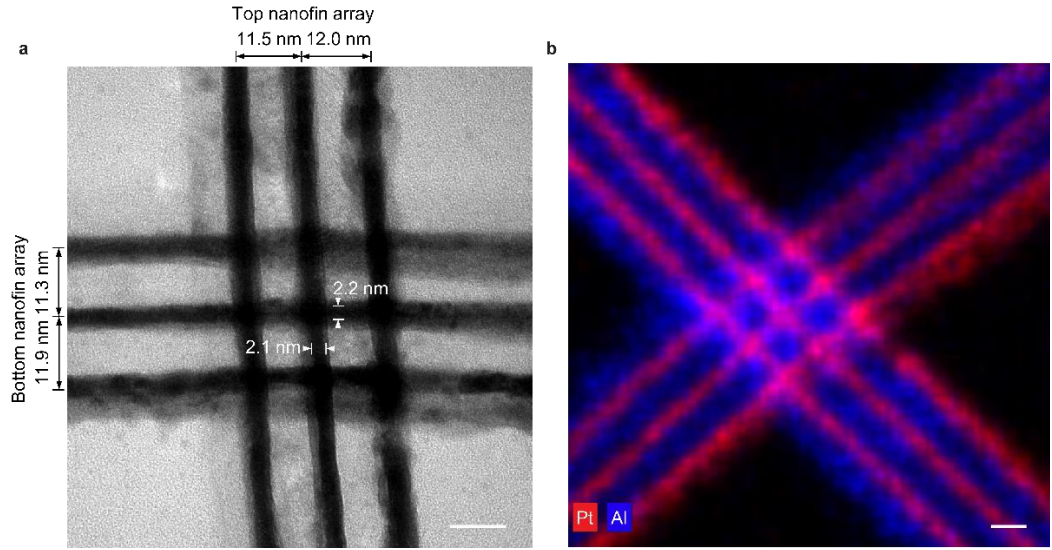

**Figure 4 | 2 nm memristor crossbar. a**. A representative TEM image of a 3 × 3 memristor crossbar with 2 × 2 nm² device area and with sub-12 nm pitch, corresponding to a packing density of 4.5 Tbits/in². **b.** Energy-dispersive X-ray spectroscopy mapping for Pt signal and Al signal to identify of the nanofins and isolation layers. All scale bars stand for 10 nm.

The 2 nm memristors in the array exhibited typical nonvolatile bipolar resistance switching between binary states (0.14 GΩ for ON state and 63.6 GΩ for OFF state, read at -1.5 V) (Fig. 5a). Several important features were observed during our electrical characterization. First, the required programming current is lower than 50 nA, leading to a peak programming power of $2.3 \times 10^{-7}$ W or 230 nW that is two orders of magnitude lower than other reported $HfO_2$ devices [11]. The read operation power is much lower, and can be further reduced by using fast electric pulses. Second, only a negative voltage on the top electrode switched the device, while the current is extremely low when a positive bias is applied. The self-rectifying property (average rectifying ratio $R_{-1.5V}/R_{1.5V} = 1.06 \times 10^3$) is critical for building large crossbar arrays without introducing external selector devices to suppress the sneak current [27,34]. It is worth noting that self-rectification is probably an ideal solution to sneak path issues at such extremely scaled size for which transistors or even most selectors are not an option for access devices any more.

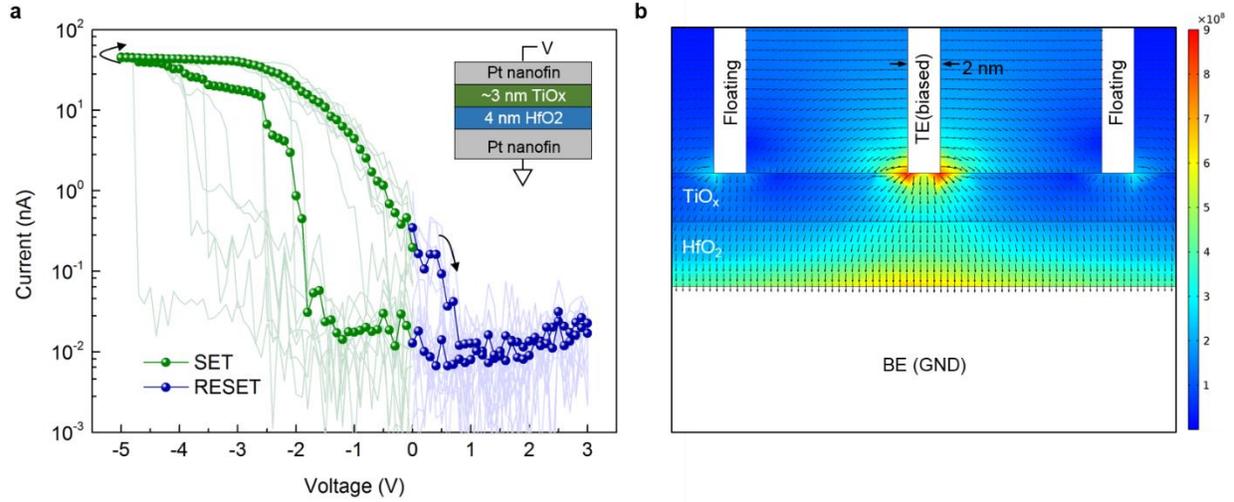

**Figure 5 | Negligible crosstalk in the 2 nm memristor crossbars. a**. A typical I-V curve of a 2 nm Pt/TiO$_x$/HfO$_2$/Pt memristor in the array. The voltage is swept from -5 to +2 V. The device exhibits bipolar non-volatile switching behavior with high dynamic range and low power consumption. The 46 nA operation current leads to a current density of $1.15 \times 10^6$ A/cm$^2$, within the limit of electro-migration [13,37]. The inset shows the device stacking geometry of the memristor. During IV measurement, top electrode is biased with the bottom electrode grounded as illustrated in the inset. **b.** Simulated electric filed distribution in 2 nm memristor crossbars with the center cell selected are plotted for the cross-section view along BE direction. The red regions indicate the strongest electric filed sensed in the junction and are used for identify the size of the effective switching junction.

The self-rectifying switching behavior of the 2 nm memristors is a result of the asymmetry of the oxide/electrode interfaces. While the HfO$_x$/Pt is intact during switching, electric field at the Pt/TiO$_x$ interface modulates the formation and rupture of a conductive channel within the TiO$_x$ layer [35]. A negative voltage on the top electrode drives oxygen vacancies towards the Pt/TiO$_x$ interface, leading to the low resistance state of the device. A positive voltage, on the other hand, repels the oxygen vacancies away from that interface, resulting in a high resistance state by recovering the electronic barrier at the interface [35]. Since the junction size is only 2 nm, there are limited number of ions in the conductive channel. Accordingly, we observed a low switching current corresponding to a current density comparable to normal resistance switching at larger sizes [11,35], i.e., $1.15 \times 10^6$ A/cm$^2$, in contrast to the four orders of magnitude higher current density reported in previous sub-10 nm devices [13]. However, a thinner channel is more susceptible to possible re-oxidation by surrounding oxygen ion/atoms and may have a shorter life time [15,36], consistent with our observation (*Supplementary Fig. 18*). Further optimization in device performance are required such as reducing oxygen concentration in the matrix through material engineering.

The crosstalk between the highly packed memristor is negligible. This is resulted from the low operational current and highly localized electric field for the 2 nm devices, which is confirmed with the similar nonlinear and low current switching behavior for all devices in the entire array (*Supplementary Fig. 19*). We simulated the electric field distribution within the memristor crossbars using a COMSOL Multiphysics model (details in *Supplementary Fig. 20*). With the selected device biased under 5 V and all others floating, the electric field is concentrated within a ~2 nm range from the electrode edge (Fig. 4b). The electric field reduces from its maximum ~$9\times10^8$ V/m at the selected electrode to nearly one order lower at the neighboring cells, which has negligible effect on the unselected devices. We also simulated the current induced heating within the crossbars. With the low current operation mode, the temperature at the switching center is 301 K (assuming ambient temperature of 300 K), and the temperature raise at the neighboring unselected cells is only less than 1 K (*Supplementary Fig. 21*). Such a cold switching mode supports a highly stable switching environment with nearly zero thermal crosstalk happening even in higher packing density, which breaks a great chance to implement memory with unprecedentedly high density (*Supplementary Fig. 21*) [28].

In summary, we have experimentally constructed a working memristor array with 2 nm feature size and 4.5 Tbit/inch$^2$ packing density for the first time using a newly developed nanofin electrode technology that breaks the series resistance limit associated with the scaling of electronic devices. The fabrication approach is highly reliable (100% yield in our case), scalable (down to 1 nm domain), and flexible as the critical dimension and the pitch of nanofins can be modulated by tuning the deposition thickness of metal and isolation layer. The 2 nm device exhibits bipolar resistance switching with an operational current of 46 nA and peak programming power of only 0.23 µW, demonstrating that scaling is an effective approach to improve power efficiency for beyond-CMOS devices. This work proves the extreme scalability of memristors experimentally and opens an avenue to dense memristor arrays with low power consumption for both memory and unconventional computing applications. The fabrication technology also provides a robust approach to build other types of artificial nanostructures with excellent dimension controllability for other applications.

**Methods**

**Substrate preparation.** Rectangular trenches (680 µm long, 70 µm wide and ~5 µm deep) were made into a Si wafer by using photolithography (5 micron thick KL-6008 positive photoresist, KemLab Inc.) and reactive ion etching (STS Vision 320 RIE system) (2 sccm $SF_6$ and 20 sccm $CHF_3$, pressure 15 mTorr). After photoresist stripping and Piranha cleaning, the 650 nm thick thermal oxide was grown at 1100 °C in a wet oxidation process. The oxide layer was later fully removed in a diluted hydrofluoric acid solution (1:25), resulting in smooth vertical sidewalls in silicon, which was then passivated by a 400 nm thick thermal oxide with wet oxidation.

**Nanofins formation.** A 2 nm thick Ge and 2 nm Pt thin films were sequentially evaporated onto the vertical $SiO_2$ sidewalls at an 85 degree oblique angle. A pair of 180 nm thick Cu contact structures were constructed with direct contact to the Pt film by using photolithography and magnetron sputtering followed by liftoff in acetone. A blanket layer of 5.5 nm thick $Al_2O_3$ was then coated using ALD at a substrate temperature of 250 °C. The aforementioned processes were repeated to form the nanofin array. The distance between the Cu contacts was shorter than the previous layer to avoid overlapping. A 5 µm thick PECVD $SiO_2$ was deposited on the sample surface, followed by chemical-mechanical polishing that led to ~3 µm tall nanofins.

**Memristor crossbar fabrication.** ~2 nm thick Ti was evaporated onto one nanofins chip and then oxidized by using oxygen plasma treatment at 0.1 Torr. A ~4 nm thick $HfO_2$ was deposited by using ALD (pre-cursor of tetrakis(dimethylamido)hafnium/water) at 250 °C on top of the $TiO_x$. After surface treatment with oxygen plasma to promote adhesion, this chip directly bonded to another nanofin chip in a homemade bonding system and a customized alignment method (See Supplementary Information for more detailed information), concluding the crossbar fabrication.

**Electrical measurements.** To access the embedded contact pads, RIE was used to strip the back-side thermal oxide, followed by deep reactive ion etching (SPTS Rapier DRIE system) to thin the top silicon wafer to several micrometers thick. At this thickness, the microscale structures near the bonded interface is visible for alignment in the following procedures. 40 µm by 40 µm windows were opened to reach the Cu pads for both top and bottom electrodes by photolithography and RIE. The I-V characterization was performed by using a Keithley 4200 SCS semiconductor parameter analyzer.

**TEM imaging.** To prepare the lamella sample for TEM imaging, the sample was first thinned in a similar fashion as described above for electrical measurement. Then a small slice that contained the crossbar structures were extracted from the thinned chip and attached to Cu grids (Omniprobe lift-out grids from Ted Pella, Inc.) by using focused ion beam (FEI Helios 660 from Harvard University). The grids were then rotated/tilted to a proper angle so that the bonding interface could be aligned approximately parallel with the ion beam. Further thinning of the sample was carried out by following a standard procedure of TEM sample preparation and the final thickness was approximately 90 nm.

**Data availability.** The data that support the plots within this paper and other finding of this study are available from the corresponding author upon reasonable request.

**Acknowledgements** This work was supported by the U. S. National Science Foundation (NSF) (ECCS-1253073). Part of the device fabrication work was conducted in the clean room of Center for Hierarchical Manufacturing (CHM), an NSF Nanoscale Science and Engineering Center (NSEC) located at the University of Massachusetts Amherst. The TEM work used resources of the Center for Functional Nanomaterials, which is a US DOE Office of Science Facility, at Brookhaven National Laboratory under Contract No. DE-SC0012704.

**Author contributions** Q. X. conceived and designed the experiments. S.P. fabricated and measured the circuits. C.L., W. X. and H.X. conducted the FIB and TEM characterization. S.P., H.J., J.J.Y., Q.X. analyzed the data. Q.X. and S.P. wrote the manuscript. All authors commented and approved the manuscript.

**Competing financial interest** The authors declare that they have no competing financial or non-financial interests.